\newcommand \be{\begin{equation}}
\newcommand \ba{\begin{eqnarray}}
\newcommand \ea{\end{eqnarray}}
\newcommand \ee{\end{equation}}

\documentclass[twocolumn,showpacs,aps,prl]{revtex4-1}
\usepackage{float}
\usepackage{color}
\usepackage{epsfig}
\begin{document}
%\preprint{AIP/123-APL}

\title{Temperature controlled L\'evy flights of minority carriers in photoexcited bulk $n$-InP}
\author{Arsen V. Subashiev}
\email[Electronic mail: ]{subashiev@ece.sunysb.edu}
\author{Oleg Semyonov}
\author{Zhichao Chen}
\author{Serge Luryi}
\affiliation{Department of Electrical and Computer Engineering,
State University of New York at Stony Brook, Stony Brook, NY,
11794-2350}

\begin{abstract}

We study the spatial distribution of minority carriers arising from their anomalous photon-assisted diffusion upon photo-excitation at an edge of $n$-InP slab for temperatures ranging from 300K to 78K. The experiment provides a realization of the ``L\'evy flight'' random walk of holes, in which the L\'evy distribution index $\gamma$ is controlled by the temperature. We show that the variation $\gamma (T)$ is close to that predicted earlier on the basis of the assumed quasi-equilibrium (van Roosbroek-Shockley) intrinsic emission spectrum, $\gamma=1-\Delta /kT$, where $\Delta (T)$ is the Urbach tailing parameter of the absorption spectra. The decreasing $\gamma$ at lower temperatures results in a giant enhancement in the spread of holes -- over distances exceeding 1 cm from the region of photo-excitation.

\end{abstract}

\maketitle
%\section{Intro}
Random walk of the L\'evy flight kind arises in many diverse fields, including hydrodynamics \cite{hydro}, biology \cite{biol}, financial markets \cite{fin}, earth science \cite{earth}, etc. It is very important whenever the photon transport is mediated by multiple photon absorption-reemission processes in gases \cite{phot}, in particular in astrophysics \cite{Ivanov}. Recently, the L\'evy flight was discovered in photon-assisted transport of minority carriers in semiconductors \cite{LuryiPRB}. 

In the normal transport process, the stationary distribution of minority carriers produced by optical excitation in semiconductors decays exponentially from the excitation area and is characterized by a micron-scale diffusion length $l$. However, in moderately-doped  direct-gap semiconductors with high radiative efficiency, the minority-carrier transport is strongly modified by the ``photon recycling'' process (repeated radiative recombination and reabsorption of emerging photons). The distribution of steps in this photon-mediated random walk of minority carriers is defined by the photon reabsorption probability \cite{LuryiPRB} and the single-step probability distribution ${\cal P}(x)$ can be calculated from the interband absorption and radiative recombination spectra \cite{Luryi}. 

At the red edge of the spectra in the Urbach-tail region the reabsorption length becomes anomalously large. As a result, ${\cal P}(x)$ asymptotically obeys the power law \cite{Semyon1}
\be
{\cal P}(x) \sim  1/x^{1+\gamma}
\label{StepDistr}
\ee
with $0 \le \gamma \le 1$. Since the distribution (\ref{StepDistr}) is ``heavy-tailed,'' 
the diffusivity [conventionally defined through the second moment of ${\cal P}(x)$] diverges and the random walk is governed by rare but large steps. This kind of transport is known as the L\'evy flight \cite{Bouchaud,Shlesinger,Metzler}. 

The resultant stationary spatial distribution is limited by: (i) loss of photons in the free-carrier absorption (FCA) and (ii) loss of minority carriers in nonradiative recombination. In $n$-doped InP the room-temperature FCA coefficient is $\alpha_{\rm FCA}$ [cm$^{-1}$] $\approx 1.3\times 10^{ -18} N_{\rm D}$ [cm$^{-3}$] and decreases further at lower temperatures \cite{AOS,Semyon2}. Thus, for moderately doped samples this mechanism of loss is ineffective. The relative rates of recombination are characterized by the radiative efficiency $\eta = \tau_{\rm nr}/(\tau_{\rm nr}+\tau_{\rm rad})$ (where $\tau_{\rm rad}^{-1}$ and $\tau_{\rm nr}^{-1}$ are, respectively, the radiative and nonradiative recombination rates). The average number of steps is given by the recycling factor $\Phi = \eta /(1- \eta )$, which is typically $\Phi \gg 1$. 

In the previous study \cite{LuryiPRB}, we directly observed  a power-law decay of the hole concentration, characteristic of the L\'evy flight, as a function of the distance from the excitation --- up to several millimeters ---  for differently doped samples at 300~K. Here we demonstrate that the index $\gamma$ of the L\'evy distribution can be controlled by varying temperature. The decrease of $\gamma$ at lower $T$ produces further enhancement in the hole spread, reaching centimeter-scale distances.

%\section{Exper}

The temperature dependent luminescence was studied in the same geometry as earlier \cite{LuryiPRB}, using $n$-InP slab \cite{NIKKO} of size 20 mm by 7 mm and thickness $d$=350 $\mu$m, corresponding to $x$, $y$, and $z$ directions, respectively. The moderate doping level $N_{\rm D}= 3 \times 10^{17}$ cm$^{-3}$ gives optimum radiative efficiency at room temperature \cite{Luryi}. A 808-nm laser beam was focused on the 7-mm edge by a cylindrical lens producing a uniform excitation along the edge. The excitation energy $E =1.53$ eV was well above InP bandgap ($E_g \approx 1.35$ eV) and generated holes only in a submicron layer near the edge. We registered the intensity and spectra of luminescence emitted from the broadside as a function of the distance $x$ from the photo-excited edge.

%%%%%%%%%-----------------------------
\begin{figure}[t,b]
\epsfig{figure=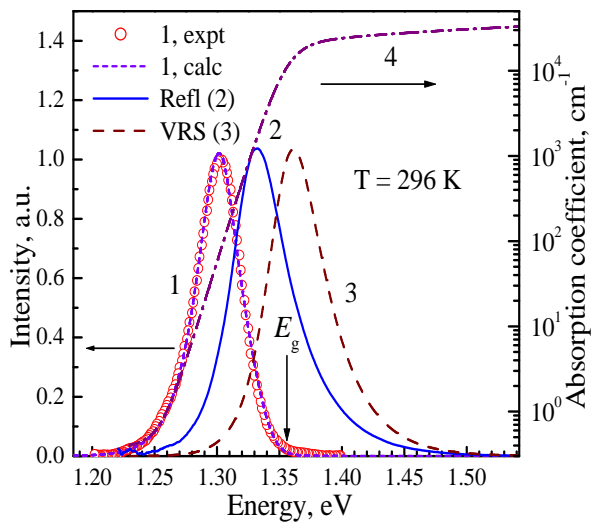,width=7.8cm,height=7.0cm} 
\caption[]
{(Color online) Luminescence spectra of $n$-doped InP sample. Photo-luminescence spectra (1) for edge excitation and front-side observation; experimental data (circles) are practically indistinguishable from the calculation (dashed line) that allows for spectral filtering. Reflection luminescence spectra from the front side is shown by the solid line (2). The van Roosbroek-Shockley intrinsic emission spectrum, Eq.(\ref{VRS}) is shown by the dashed line (3). The dash-dotted line (4) corresponds to the absorption coefficient.} \label{SpecCompare}
\end{figure}
%%%%%%%%------------------------------

The luminescence spectra retain identical shape in the whole range $x$. This important observation remains valid for all temperatures studied. The edge-excited photoluminescence spectrum for $T=296$ K, observed at $x$=1 mm, is shown in Fig. \ref{SpecCompare} together with the calculated spectrum. Also shown is the luminescence spectrum, observed from the broadside in reflection geometry \cite{Luryi,Semyon1}. We note that, compared to the reflection spectrum, the edge-excitation spectrum is noticeably shifted to the red side, indicating substantial filtering \cite{LuryiPRB}.

The absorption spectrum $\alpha(E)$ of the sample is shown in Fig. \ref{SpecCompare} by the dash-dotted line. In the whole range of temperatures studied, the red wing of $\alpha(E)$ exhibits an exponential Urbach behavior \cite{AOS},
\be
\alpha_i(E)=\alpha_0 \exp\left [\frac{E-E_g}{\Delta(T)}\right]~, \hspace{1cm} E<E_g
\label{Urb}
\ee
extending down to $\alpha(E) \approx \alpha_{\rm FCA}$. At $E\ge E_g$, the  Urbach exponent saturates. Note that that the luminescence spectra observed with the edge excitation are fully in the Urbach tail region of the absorption spectrum  --- in contrast with the intrinsic emission spectrum. 

Assuming a quasi-equilibrium energy distribution of holes, the intrinsic emission spectrum is given by the well-known van Roosbroek-Shockley (VRS) relation \cite{VRSh},
\be
S_{VRS}(E) \sim \alpha_i(E) E^3  \exp \left(-\frac {E}{kT}\right)~.
\label{VRS}
\ee
The VRS spectrum is also shown in Fig. \ref{SpecCompare}. It has a maximum above $E_g$. Owing to the rapid energy relaxation of holes, the  intrinsic emission spectrum in $n$-InP is well described by Eq. (\ref{VRS}), both at room and lower temperatures, with a noticeable deviation in shape only at $T \le$ 78~K. The observed red-shifted spectrum $S(E)$ is shaped by reabsorption on its way out of the sample, $S(E)=S_{VRS}\times F(E)$, where $F(E)$ is the spectral filtering function $F(E)=F_1(E)~T(E)$ that depends on the hole distribution $p(z)$ across the wafer and is affected by reflections from the sample surfaces. The one-pass filtering function $F_1(E)$ is given by \cite{Semyon1,Luryi}
\be
F_1(E)=\int_0^d p(z)\exp[-\alpha(E)z]dz~.
\label{Filt}
\ee
%where $z$ is directed normally from the broadside into the slab. 
The factor $T(E)=[1-R\exp(-\alpha(E) d)]^{-1}$ accounts for the multiple surface reflections; $R\approx 0.33$. Note that due to the high index contrast the radiation escape cone is narrow, i.e. the outgoing radiation propagates close to the normal direction to the surface. Effects of multiple reflections are noticeable only in the far red wing of the spectrum. For the edge excitation, the hole distribution across the sample is nearly homogeneous. Therefore, the typical radiation escape distance in this geometry is much larger and produces considerably larger red shift of the spectrum than for the reflection geometry, where holes are generated near the surface. The filtered spectra, calculated using the VRS relation, are shown in Fig. \ref{SpecCompare} by the dashed line (1), demonstrating excellent agreement with the experiment \cite{Excuse}.      

The luminescence intensity distributions $I(x)$ were obtained by scanning the CCD image along a line parallel to $x$, as in the earlier experiment \cite{LuryiPRB}. The proportionality $p(x)\propto I(x)$ is supported by the good agreement between the calculated and observed luminescence spectra for edge excitation. 

\begin{figure}[t]
\epsfig{figure=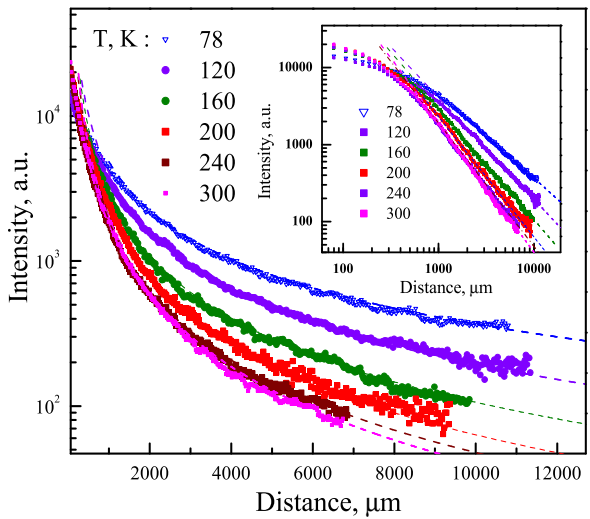,width=7.8cm,height=6.9cm} \caption[]
{(Color online) Distribution of the luminescence intensity $I(x)$ and the hole concentration $p(x) \propto I(x)$ for different temperatures $T$. Dashed lines correspond to the power-law (\ref{StepDistr}) fitting with an index $\gamma$. The inset shows the same distributions in log-log scale.   } \label{SpecvsX}
\end{figure}

The resulting distributions for temperatures ranging from 300~K to 78~K are shown in Fig. \ref{SpecvsX}.  One can see a huge enhancement of the hole spread that extends over 1 cm at temperatures below 200 K.

The power law $p(x) \sim 1/x^{1+\gamma}$ is clearly observed for all temperatures at distances $x>0.5$ mm. This is best seen on the log-log scale in the inset of Fig.~ \ref{SpecvsX}. The power law is in clear contrast to an exponential decay $p(x) \sim \exp (-x/l)$  expected for a normal diffusion of holes, even accounting for any photon-assisted enhancement of the diffusion length $l$.
%%%%%%%%%%%----------
%%%%%%%%-------------------------------

The observed hole distribution $p(x)$ for all temperatures follows the calculated distribution for the L\'evy flight random walk governed by the single-step probability (\ref{StepDistr}). Details of this calculation have been extensively discussed \cite{Ivanov,Bib,Holstein,Luryi}.  The ${\cal P}(x)$ itself is evaluated as the photon reabsorption probability \cite{LuryiPRB,Luryi}; it is fully determined by the absorption and the emission spectra. It can be calculated numerically in the full range of distances. At large distances ($x \gg \alpha_0^{-1}, l$) a tangible contribution to ${\cal P}(x)$ results only from the exponentially varying Urbach part of the absorption spectra. This region also correspond to the exponentially decaying red wing of the intrinsic emission spectrum. Assuming the intrinsic spectrum in the VRS form, we find \cite{Semyon1,Luryi} an analytic expression for index $\gamma$ of the single-step distribution  
\be \gamma_{VRS} = 1-\frac{\Delta}{kT}~. \label{gammaVRS}
\ee 

An analytic expression for the hole distribution $p(x)$ can be derived \cite{SL12} in the so-called ``longest flight'' approximation \cite{Ivanov,BigJump,SL12} and is of the form
\be
p(x)=\frac{c x_f^\gamma}{x^{1+\gamma}[1+(x_f/x)^\gamma]^{2}}~,
\label{interp}
\ee
where $c$ is a normalization constant.
Distribution (\ref{interp}) provides a good approximation to the exact solution in the entire range of $x$. Asymptotically (at $x \gg x_f$) it reproduces the single-step probability (\ref{StepDistr}), exhibiting a power-law decay with index $\gamma$. At short distances, $x \ll x_f$, it gives a weaker decay, $p(x)\propto x^{\gamma-1}$, in agreement with the exact solution \cite{LuryiPRB,SL12}. The distance $x_f$ gives an estimate of the spread of the excitation ``front'' of $p(x)$ --- beyond which the holes appear predominantly in one step. The transition between the short- and long-distance asymptotics is clearly seen for all $T$ in the inset to Fig. \ref{SpecvsX}, with $x_f$ being the point of maximum curvature in the log-log plots. 

%%%%%%%%%-----------------------------
\begin{figure}[t,b]
\epsfig{figure=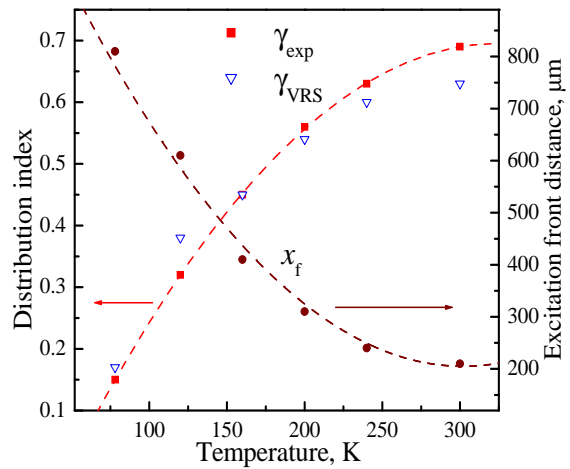,width=7.5cm,height=6.2cm} 
\caption[]
{(Color online) Temperature dependence of the parameters of L\'evy distribution. Index $\gamma$ shown by squares is estimated from the long-distance asymptotics of the measured $p(x)$; also shown (triangles) are the theoretical values (\ref{gammaVRS}). The excitation front distance $x_f$ is estimated from maximum curvature in the log-log plots of Fig. \ref{SpecvsX}. Lines are guides to the eye.} \label{IndXf}
\end{figure}
%%%%%%%%------------------------------

We have estimated  $\gamma$  through the slope of log-log variation at far distances and also $x_f$ via the position of maximum curvature of $p(x)$.  The temperature variation of obtained values of $\gamma$ and $x_{f}$ is shown in Fig. \ref{IndXf}. Also shown are theoretical values of the index $\gamma_{VRS}$ calculated (\ref{gammaVRS}) by assuming the intrinsic emission spectrum in the VRS form (\ref{VRS}). We see that the estimation of $\gamma$ through the Urbach tailing parameter gives a good agreement with experiment. 

Position of the excitation front can also be estimated \cite{LuryiPRB,Luryi,Semyon1} through the recycling factor, viz. $x_f=x_0\Phi^{1/\gamma}$, where $x_0\approx 0.2$ $\mu$m, is the length scaling parameter of ${\cal P}(x)$. The observed huge increase of the excitation front distances, $x_f\gg x_0$ up to the mm-range at low temperatures is another hallmark of L\'evy  flight, which cannot be attributed to any other mechanism of hole transport.  We note however, that for $\Phi \approx 100$ and observed $\gamma(T)$ one should expect even higher increase of  $x_f$. Therefore, our experiments suggest a {\it decreasing} recycling factor at lower $T$. Apparently, the well-known increase of the radiative recombination rate at low $T$ is accompanied by an even faster growth rate of non-radiative recombination, as found for several recombination models with vanishing activation barrier \cite{yass,rosen,blaikh}.  

From a practical perspective, the anomalous transport of minority carriers should be important in all semiconductor devices with high radiative efficiency \cite{Luryi}. However, details of the transport and the temperature effects depend on the absorption spectrum in the tailing region and are not universal, e.g. often deviate from the Urbach shape. Thus, some semiconductors feature  Gaussian tails \cite{Eliseev}. We have calculated the index $\gamma$ for Gaussian tails and found it to {\it increase} with decreasing temperature, remaining  within a narrow range ($0.85 < \gamma <1.05$) close to unity for a wide temperature variation. Such semiconductors may serve as an experimental model for studying the transition from L\'evy flight ($\gamma < 1$) to L\'evy walk ($1 < \gamma < 2$). To our knowledge, this transition has been considered only as a notional possibility  \cite{Metzler}.

In conclusion, we studied the stationary hole distribution $p(x)$ produced by photo-excitation at an edge of $n$-InP slab and observed by broadsided luminescence at temperatures ranging from 300 K to 78~K. We discovered a giant increase of the hole spread in the sample over distances $x$ exceeding 1 cm from the photo-excited edge. A power-law decline of the luminescence intensity characteristic of L\'evy flight kinetics is observed with no change in the spectral shape. Our experiment provides a realization of the L\'evy flight of holes, in which the L\'evy distribution index $\gamma$ is controlled by the temperature and can be continuously lowered from $\gamma\approx 0.7$ to $\gamma \le 0.3$ in a regular and well-understood manner. Such a control has never been available for formerly studied L\'evy processes.

This work was supported by the Domestic Nuclear Detection Office, by the Defense Threat Reduction Agency (basic research program), and by the Center for Advanced Sensor Technology at Stony Brook.


\begin{thebibliography}{99}
\bibitem{Excuse} Unfortunately, this detailed comparison cannot be extended to lower temperatures, since the absorption spectra are not known in the whole energy range of interest.
\end{thebibliography}
\end{document}